\documentclass[11pt,a4paper]{article}

\usepackage{graphics}
\usepackage{graphicx}
\usepackage{epsfig}
\usepackage{amssymb}
\usepackage{subfigure}
\usepackage{bm} 
\usepackage{mathrsfs}
\usepackage{color}

\usepackage{authblk}

\title{Continuous Forest Fire Propagation in a Local Small World Network Model}
\author[1]{F. Aguayo}
\author[2]{A. Fuentes}
\author[3]{J.-P. Clerc}
\author[3]{B. Porterie}

\affil[1]{School of Mathematical Sciences, 
University of Nottingham, 
University Park, Nottingham, NG7 2RD, UK}

\affil[2] {
Universidad T\'ecnica Federico Santa Mar\'ia, 
Departamento de Industrias, Av. Espa\~na 1680, Valpara\'iso, Chile}

\affil[3] {
Aix-Marseille Universit\'e, IUSTI/UMR CNRS 7343, 5 rue E. Fermi, 
13453 Marseille Cedex 13, France}

\begin{document}
\maketitle
\begin{abstract}
This paper presents the development of a new continuous forest fire
model implemented as a weighted local small-world network
approach. This new approach was designed to simulate fire patterns in
real, heterogeneous landscapes. The wildland fire spread is simulated
on a square lattice in which each cell represents an area of the
land's surface. The interaction between burning and non-burning cells,
in the present work induced by flame radiation, may be extended well
beyond nearest neighbors. It depends on local conditions of topography
and vegetation types. An approach based on a solid flame model is used
to predict the radiative heat flux from the flame generated by the
burning of each site towards its neighbors. The weighting procedure
takes into account the self-degradation of the tree and the ignition
processes of a combustible cell through time. The model is tested on a
field presenting a range of slopes and with data collected from a real
wildfire scenario. The critical behavior of the spreading process is
investigated.
\end{abstract}

\section{Introduction}

The forest fire propagation models are usually categorized into
stochastic and deterministic models. Concerning the deterministic
models, the fire behavior is deduced directly from conservation
equations and imposed physical laws driving the evolution of the
system. Several sophisticated models exist in the literature. Based on
Weber's classification, it is possible to identify three types of
mathematical models according to the methods used in their
construction \cite{web}. The first contains the statistical models
which do no attempt to include specific physical mechanisms, using
only a statistical description of the fire \cite{mcar}. The second
kind of model {\color{black} are} composed by the semi-empirical
models. They are based on energy conservation but do not pretend to
distinguish the different modes of heat transfer
\cite{rothe}. Finally, the physical models aim to solve (with some
approximations) the equations governing fluid dynamics, combustion,
and heat transfer. Many multidimensional transient wildfire simulation
approaches have been developed. They are based on methods of
computational fluid dynamics including the vegetative fuel and
fire-atmosphere interaction which allows variation of the degrees of
complexity of the simulation\cite{mell1,lin,sero,mor}. However,
{\color{black} some} disadvantages appear with the latter approaches,
where the {\color{black} reults tends to be gratly influenced by} the
discretization technique. Moreover, the filters, constituted by both
physics-based concepts and discretization techniques, result in
multidimensional parameters which {\color{black} are not always
  directly related to the physically meaning} properties of the forest
fire. The higher computational cost required to solve the equations
also needs to be considered.

On the other hand, heterogeneous conditions of weather, fuel, and
topography are generally encountered during the propagation of large
fires \cite{dimi}. Irregular shapes contours and fractal post-fire
patterns, as revealed by satellite maps \cite{cal}, suggest that
stochastic models are good candidates for studying the erratic
behavior of large fires. From a stochastic point of view, the forest
fire spread has been modeled using regular networks, as well as
cellular automata in order to include site weights
\cite{Alb86,Tep97,Bal92,Bar97}. However, there are some evidence that
networks with only local contacts do not mimic appropriately real
fires \cite{gen,new}.

Also, the wildland fire can be regarded as an example of percolation
phenomena. In its simplest form, a mesh with an occupancy probability
$p$ is used to represent the forest. The fire is introduced in this
mesh allowing it to ``infect'' the nearest neighbors of a tree already
on fire. In many physical systems, this classical percolation picture
is modified in many various ways, changing the spread mechanism, the
way the mesh is designed, the range of interaction, etc. For example,
in the case of forest fires, it is possible to find from medium to
long range interactions due to the nature of the radiative heat
transfer and/or spotting effects. These types of interactions will
affect the way the mesh is constructed: the links and their
``weights'' can be changed.

Based on the characteristics mentioned above, the stochastic model
used in this study was developed by Porterie et
al. \cite{porterie07,Zek05}.  It is a variant of the so called local
small-world network model ($lswn$) \cite{watt}. It considers flame
radiation as the mechanism of fuel preheating and spotting effect,
thus including long-range interactions, well beyond the nearest
neighbors. A weighting procedure on combustible cells is used to take
into account latency and flaming persistence of burning cells. It is
based on the knowledge of two characteristic times, namely the time
required for a site to achieve complete combustion, $t_c$, and the
time required for thermal degradation before ignition, $t_{td}$
\cite{Por08}.

The model proposed here describes a deterministic evolution of the
fire spread {\color{black} over time}. It allows the incorporation of
the advantages of $lswn$ model presented by Porterie et
al. \cite{porterie07} by introducing a new description of the dynamic
process into the $lswn$ model. The fire spread starting from the
interaction between linked sites is accounted giving a
phenomenological description of the interaction between neighboring
sites (trees) in order to construct the whole spreading process. This
phenomenology is represented via a system of coupled first order
differential equations for the normalized mass at each site.

\section{Formulation (continuous fire spread model)} 

A continuous fire spread model scheme was adopted instead of a
discrete one as a way to perform a more systematic study of the
involved parameters.  Discrete models gives robust framework to
understand the propagation phenomena, but when more effects are to be
taken into account, a continuous representation seems more appropriate
to represent all the underlying physics.

In particular, we focus our attention in the effect of the terrain
slope in the spread of a forest fire. As discussed in \cite{Zek05} and
\cite{Por08}, the types of interactions are essentially non-local,
leading us to adopt a \emph{lswn} approach to take into account the
interaction further away than first neighbors.  In the isotropic case,
\emph{i.e.} with no wind or slope, this construction leads to a system
that can be regarded as a rank-l network, which would enable to use
the well known renormalization techniques \cite{New99}.  In practice,
the randomness of a real set-up, {\color{black} prevents} us to do so.
The effects of the slope will break the symmetry of the network, and
even turning it into a directed graph because the propagation may be
possible only in certain directions. This directionality will be
local, highly dependant in the place the site occupies in the
neighborhood. Furthermore, not only the topology will be affected, but
also its links weight.  The experience confirms that the anisotropy
induced by these agents also affects the dynamic properties of the
system, as the rate of spread of the fire front \cite{porterie07}.

Finally, the ``microscopic'' (interaction between trees) 
dynamic will also depend on the history 
of each site and not only on its current state, in the sense the
ability of a site to infect its neighbors will change with time.
For the above mentioned reasons, we propose a continuous spreading
model over a \emph{lswn}, which is spawned over the randomly occupied
sites of a regular lattice.

The procedure to generate the network is as follows. A given
regular lattice is filled with occupancy probability $p$, related to
the density of the medium, in our case, the density of trees in a 
forest. For the occupied sites, the linking neighbors are to be 
found, in principle all the forest could be searched, but for 
computational convenience, in this work the neighborhood is 
comprised of all the active sites within $1.5$ 
times the elemental mesh length. In Section \ref{sec:diff-eq} we 
will see that this is an adequate first
approximation. Once the network and links are established the 
dynamics is controlled by the continuous spreading model.

\subsection{Models Assumptions}

In the following, we will assume the radiative heat transfer is the 
predominant mechanism in the propagation of forest fires. Although,
convection can have important repercussions under certain
circumstances, radiation has been shown to be the most important
factor in propagation \cite{porterie07}.

The state of the site is controlled basically by the evolution of 
the amount of fuel pyrolysized. Moreover, this quantity is driven 
by different phenomena: ignition, burning rate and propagation. 
At the beginning, the decomposition of the solid fuel available 
from the tree (leaves, branches, etc.) is principally controlled 
by radiative heat transfer from a neighboring tree in flames. 
Once the solid fuel ignites, the self burning process will be 
the dominant phenomena (burning rate). Thus, we will study the 
two terms separately, each one in the absence of the other, 
expecting to describe the overall process in a reasonable way.

To adequately represent the fire propagation between linked trees,
we will start from the following set of rules:  
\begin{itemize}\itemsep0pt
\item We will focus on the normalized fuel mass $C_{i}(t)$, relative
to the initial amount of fuel in the site $i$ $m_i^0$, at time
$t$. Thus, $C_{i}(t)=m_i(t)/m_i^0$ so
$C_{i}(t=0)=1$ if $i$ is active and $0$ otherwise.
\item $C_{i}$=1 and $C_{i}$=0 are stationary states of the site if no
external agent is considered.
\item In order for a site to start the ``self-combusting'' process a
minimal amount of incident radiative heat flux is needed. 
\item The radiation absorbed by a site is proportional to the 
amount of fuel left in that site.
\item The degradation of a site is proportional to the 
energy received.
\item While burning, the radiation emitted is proportional 
to the mass loss of a site.
\end{itemize}

\subsection{Two-sites interaction}
\label{sec:diff-eq}

We state that, for a site $i$ whose neighbor $j$ is
burning,
\begin{equation}
\triangle C_{i} = \alpha E^r_j = \alpha E_j 
f(\bm{x}_i-\bm{x}_j)r 
C_{i},
\label{eq:delta-ci}
\end{equation}
where $E^r_j$ is the total amount of energy received as
radiation coming from the site $j$. This quantity is proportional to the
total energy released $E_j$, where the proportionality factor is {\color{black} composed} by the radiative energy 
portion $r$ (radiant fraction of the heat release rate), 
usually about $35\%$, and a geometric view
factor $f(\bm{x}_i - \bm{x}_j)$. The proportionality constant
$\alpha$, is related to the heat of ignition of the site $i$. These
two quantities will be discussed later. The term $C_{i}$ in the RHS of
Eq. \ref{eq:delta-ci} stands for the ability of the site to absorb
radiation which is proportional to the density of fuel left in the
site, considering it is mostly distributed in the surface of the
canopy. Should a volumetric distribution be
considered more adequate, a power of $2/3$ has to be used for the fuel density.

The total emitted energy, for a given period of time, can be
estimated using the amount of burned mass, as
\begin{equation}
E_j=c_{h}m_{j}^0\triangle C_{j},
\label{eq:ej}
\end{equation}
where $c_{h}$ is the heat of combustion. Assuming an infinitesimal period, Eq. \ref{eq:delta-ci} yields to, 
\begin{equation}
\frac{dC_{i}}{dt} = C_{i} \frac{dC_{j}}{dt}
\left( 
\alpha
f(\bm{x}_i-\bm{x}_j) 
r
c_{h} 
m_{j}^0 
\right) = C_{i} \frac{dC_{j}}{dt}W_{ij}.
\label{eq:dci}
\end{equation}
This simple term will be the responsible of the interaction between
sites. To take into account the whole neighborhood, a summation over
the index $j$ is needed.

\begin{figure}[h!]
\begin{center}
\resizebox{6.0cm}{!}{\includegraphics*{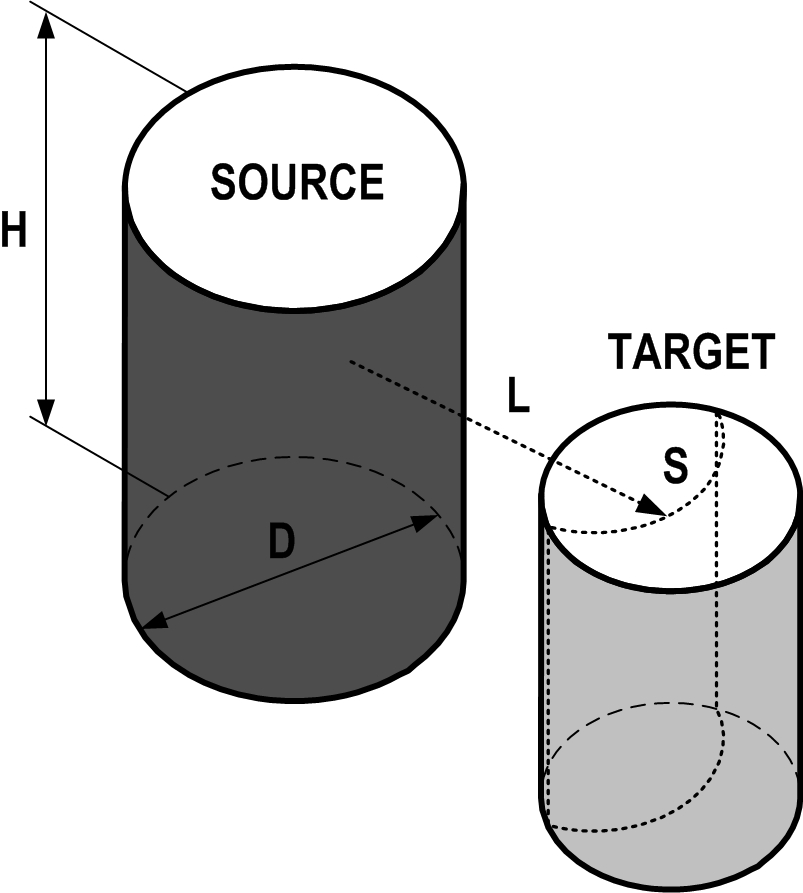}}
\caption{\label{fig:cil} Scheme of the radiation emitted by a solid flame and received by a cilyndrical target (tree).}
\end{center}
\end{figure}

In order to ignite a tree, the decomposition of the solid fuel needs 
to be initiated. 
The later becomes possible when an external source generates energy 
to heat the fuel (usually by radiation) and promotes the release 
of fuel vapors and pyrolysis gases.
Normally, the ignition of a solid fuel can be correlated to a 
critical mass flux of pyrolysis gases \cite{Atre98}.
In this study the parameter $\alpha$ can be 
readily estimated by measuring the total energy needed to ignite 
the tree ($E_i^i$), and the critical amount of fuel at which the 
ignition occurs ($C_i^{th}$),
\begin{equation}
\alpha = \frac{\ln\left(1/C_{i}^{th}\right)}{E_{i}^i}.
\end{equation}
The energy needed can be obtained via $E_{i}^i=h_{i}m^0_{i}$, where
$h_{i}$ is the heat necessary to induce the ignition 
(the onset of thermal runaway) for a given fuel. So the
coupling constant $W_{ij}$ can be written as:  

\begin{equation}
W_{ij} = 
\left( 
\frac{\ln\left(1/C_{i}^{th}\right)}{h_{i}}
f(\bm{x}_i-\bm{x}_j) 
r
c_{h} 
\frac{m_{j}^0}{m^0_{i}}
\right). 
\end{equation}
Finally, $f(\bm{x}_i-\bm{x}_j)$ is estimated considering a view factor 
between the tree with flames and the target (the neighboring tree) using 
the concept of solid flame \cite{Sull03}. In this study the source was 
considered as a cylinder
(solid flame) emitting to a target (the neighboring tree) depicted as another 
cylinder of similar or different dimensions. The radiation can be
described using analytical expressions of view factors between 
a cylinder and a infinitesimal element of surface belonging to the target.
A scheme of the set up is shown in Fig. \ref{fig:cil}. 
In the most simple scenario we consider the trees as cylinders emitting and
absorbing radiation to and from their neighbors. Accordingly to
\cite{SFPE}, for a cylindrical source and a vertical target, the view factor 
can be evaluated by:
\begin{eqnarray}
F(H) &=& \frac{1}{\pi S}
\tan^{-1}\left(\frac{h}{\sqrt{S^{2}-1}}\right) - 
\frac{h}{\pi S}
\tan^{-1}\left(\sqrt{\frac{(S-1)}{(S+1)}}\right) + \nonumber \\
& &
\frac{Ah}{\pi S\sqrt{A^{2}-1}}
\tan^{-1}\left(\sqrt{\frac{(A+1)(S-1)}{(A-1)(S+1)}}\right),
\end{eqnarray}
with:\\
$A = {(h^{2}+S^{2}+1)}/{2S}$,  $B = {(1+S^{2})}/{2S}$,  $S = {2L}/{D}$ and $h = {2H}/{D}$,\\
$L$ being the distance between the center of the cylinder to the target,
 $H$, the height of the cylinder -in the following taken to be 
about $3\;[m]$- and $D$ -about $2\;[m]$- is the cylinder diameter.

{\color{black} 
So the total incoming radiation is the integral of the flux factor over the receiving
surface $\Sigma$:
\begin{equation}
f(\bm{x}_i-\bm{x}_j) =
\int_{\Sigma} \; d^3x\; sign(H-z)\; F(\left|H-z\right|) 
+ sign(z)\; F(\left|z\right|)\\
\end{equation}
Note that $F(H)$ has been divided in to two contributions, corresponding to the 
respective portions of flame situated above and below the integration point. 
It also gives the correct contribution for the case where the point lays 
higher or lower to the flame.}

\begin{figure}[h!]
\begin{center}
\resizebox{6.7cm}{!}{\includegraphics*{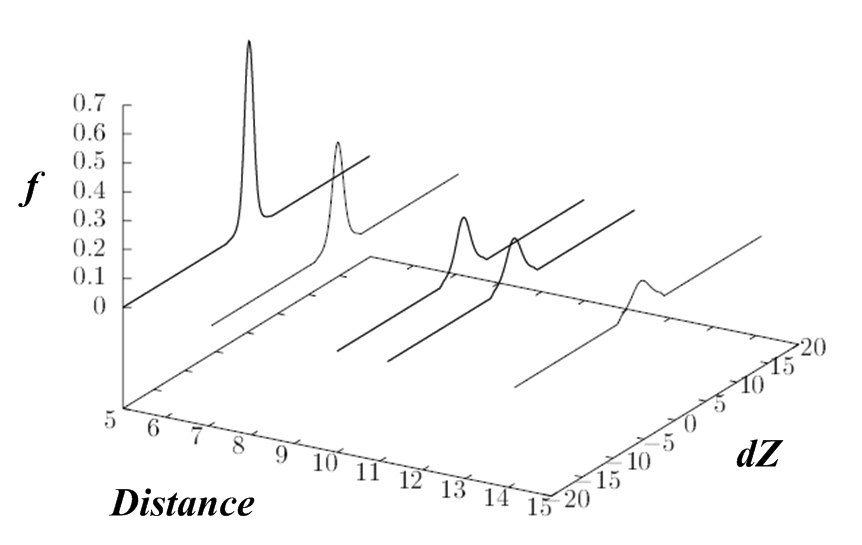}}
\caption{View factors at different distances and heights ($dZ$).
\label{fig:flux}} 
\end{center}
\end{figure}

Using this model we can estimate the radiation received by a punctual
target localized at certain distance. Now, as our target is not punctual, 
but represented by a cylinder {\color{black} it} is necessary to
integrate over its receiving area. In order to reduce computational
time and simplify the model, {\color{black} we} assume the simplified surface depicted
in Fig. \ref{fig:cil}.
Then, the arc $S$ is a portion of a circle of radius L, thus, orthogonal
to the distance. Given this, we only need to calculate the incoming radiation
at one point, and {color{black} multiplying by the} length of the portion of arc
S, namely
\begin{equation}
S=2L\left(\pi-2cos^{-1}\left(\frac{r}{2L}\right)\right),
\end{equation}
$r$ being the radius of the target ($r\sim3\;[m]$).
{\color{black} Finally we have to} integrate along the height of the target to get
the total incoming radiation. This last step is done numerically and the resulting 
radiation at different distances, as function of the 
difference of height $dz$ is shown in Fig. \ref{fig:flux}.
{\color{black} To incorporate the slope in the description, it is sufficient
to change the range of the integral over the receiving canopi, so the height
difference between sites is properly reflected.

Finally, to take into account windy scenarios, we will use a reasonable approximation
for long distances; The radiation model 
will be taken to be valid for a thin disc of height $dh$ representing an infinitesimal
slice of the flame. Thus, a tilted flame will be composed of several of those discs which
energy flux over the neightbours can be estimated as stated before. 
The flame inclination is calculated as in \cite{Alb81}.
}

Regarding Fig. \ref{fig:flux}, the quadratic decay will allow us 
to neglect the interaction between distant neighbors. This will 
allow us to keep the amount of links between each site relatively small,
and conserve the significant long range interactions. This \emph{lswn} nature
can affect largely the percolative properties of the system,
for example making the percolation threshold to fall well 
below the usual values found in first-neighbors systems.

An important final remark has to be done about this factor, 
as is here where the dependency in the surface is hidden. 
It relates the flux between two vertical surfaces (cylinders), 
so it will peak when both are vertically centered. As the 
surface representing the flame raises above the canopy, it will 
increase the propagation towards slightly higher surfaces and 
will produce a decreasing radiative flux for surfaces away from 
the center. This is the expected behavior and will drive the 
height dependency in the spreading model.


\subsection{Self degradation}

Usually, once the decomposition process is started the pyrolysis 
mass rate is mainly accelerated 
by the radiation coming from the combustion processes occurring 
in the tree. 
Experimentally, this process shows a slow initial evolution if the
canopy is ignited {\color{black} while} the moisture content {\color{black} is still} 
important. This
is followed by an acceleration of the process in the intermediate
stage and a new slow down at the end of the process, when the fuel has
been mostly consumed. For a burning tree, we propose a mass evolution
of the form:
\begin{equation} 
\frac{dC_{i}(t)}{dt}=-C_{i}(t)\left(1-C_{i}(t)\right)U_{i},
\end{equation}
which is in qualitative concordance with experiments \cite{mel09}. 
For large times,
$C_i(t)$ will decay exponentially, relating the factor $U_i$ to the
inverse of the \emph{mean-life} of the isolated burning tree, usually 
about $30s$ \cite{mel09}.

With this, the overall equation for the rate of mass over time of
every active site is:
\begin{equation} 
\frac{dC_{i}(t)}{dt} = C_{i}(t)\sum_{<j>}\frac{dC_{j}(t)}{dt}W_{ij} - 
\Theta\left[C_{i}^{th}-C_{i}(t)
\right]
C_{i}(t)
\left(1-C_{i}(t)\right)
U_{i}.
\label{eq:dc-final}
\end{equation}
In Eq. \ref{eq:dc-final}, the sum over $<j>$ has to be performed over
all the \emph{burning} neighbors of $i$, as we only account for
radiative processes. It is important to note that the second term in RHS is activated 
when the burning threshold $ C_{i}^{th}$ is reached, this is controlled by the Heaviside $\Theta$ step function.

\section{Numerical tests}

The system here described can be solved numerically with relative 
ease. Following 
we study two different scenarios to get an insight of the behavior 
of the model.
First we will see a small system of only 4 trees, where we will be 
able to analyze quantitatively the behavior of the model and the evolution of 
$C_i(t)$ for a burning 
system. Later we go to a larger system, in order to study the 
percolation like properties and its critical behavior.

In the following we will consider a mean life of a burning tree to
be $30\;[s]$, the degradation coefficient 
$\alpha=0.000005\;[J^{-1}]$, the burning threshold 
$C_{i}^{th}=0.9$, and the heat of combustion 
$c_h=12000\;[kJ/kg]$. 

\subsection{Four sites evolution}

\begin{figure}[h!]
\begin{center}
\resizebox{6.7cm}{!}{\includegraphics*{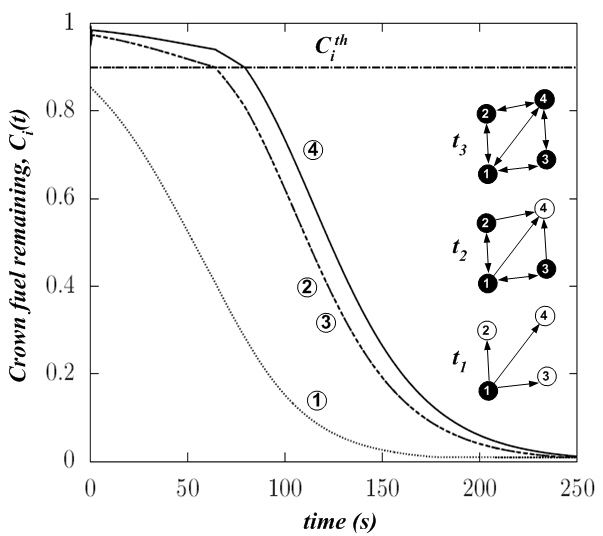}}
\caption{\label{fig:fuel} Simple interaction example of four 
trees in and flat landscape. The curves shows the fuel available 
in each site in function of time and the schematic inside the 
graph represent the interaction at different times.} 
\end{center}
\end{figure}

Once the parameters discussed in {\color{black} the} previous section
are fixed and the numerical simulation {\color{black} is} implemented,
{\color{black} it} is interesting to look the simple interaction
between four trees symmetrically distributed (see the schematics in
the Fig. \ref{fig:fuel}). The simulation was started burning only site
1. At $t_1$ only the site 1 release energy causing the degradation of
the site 2, 3 and 4.  Once the threshold of ignition $C_{th}$ is
{\color{black} reached} at $t_2$, both sites 2 and 3 are ignited
simultaneously. The evolution for the sites 2 and 3 are identical due
to the symmetry and only three curves are observed in the
graph. Regarding the evolution of each curve, there is an evident
change in the slope when any of them reaches the ignition threshold
$C_{th}$. At $t_3$ the site 4 is burning and the self burning process
accelerates the degradation. Clearly these processes of self-burning
contributes to increase the slope for the fuel mass losses evolutions
of the neighboring trees. For larger times, the mass decays
exponentially and as its derivative also goes to zero, the effect on
the neighbors also decreases. Finally, it is important to note that
the evolution of the fuel reaming in the tree along the time, exhibits
the typical behavior observed in experimental results \cite{mel09}.

%

\subsection{Front evolution}

\begin{figure}[!ht]
\begin{center}
\begin{tabular}[t]{c}
\resizebox{6.7cm}{!}{\includegraphics{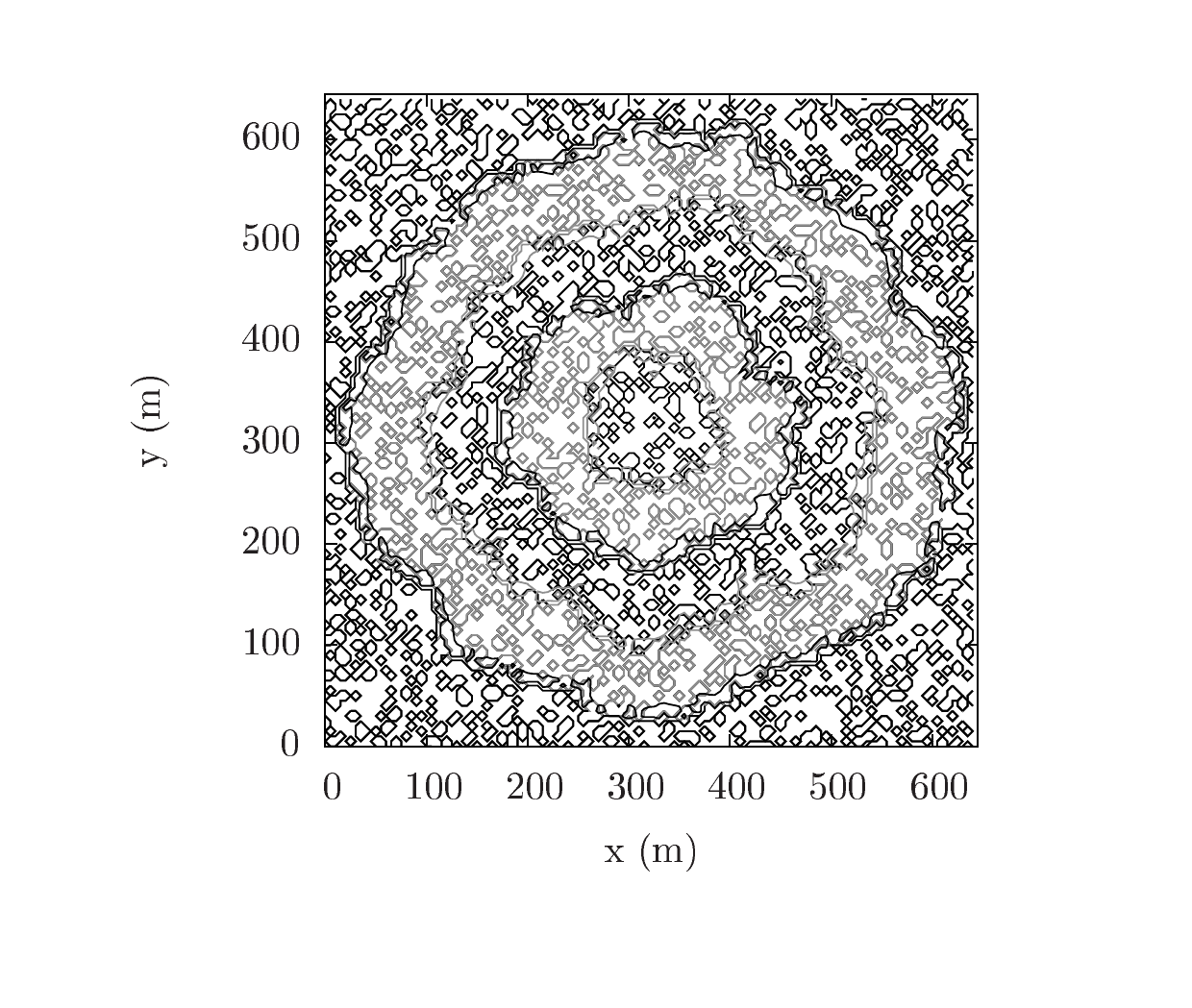}} 	\\
(a)  \\
\resizebox{6.7cm}{!}{\includegraphics*{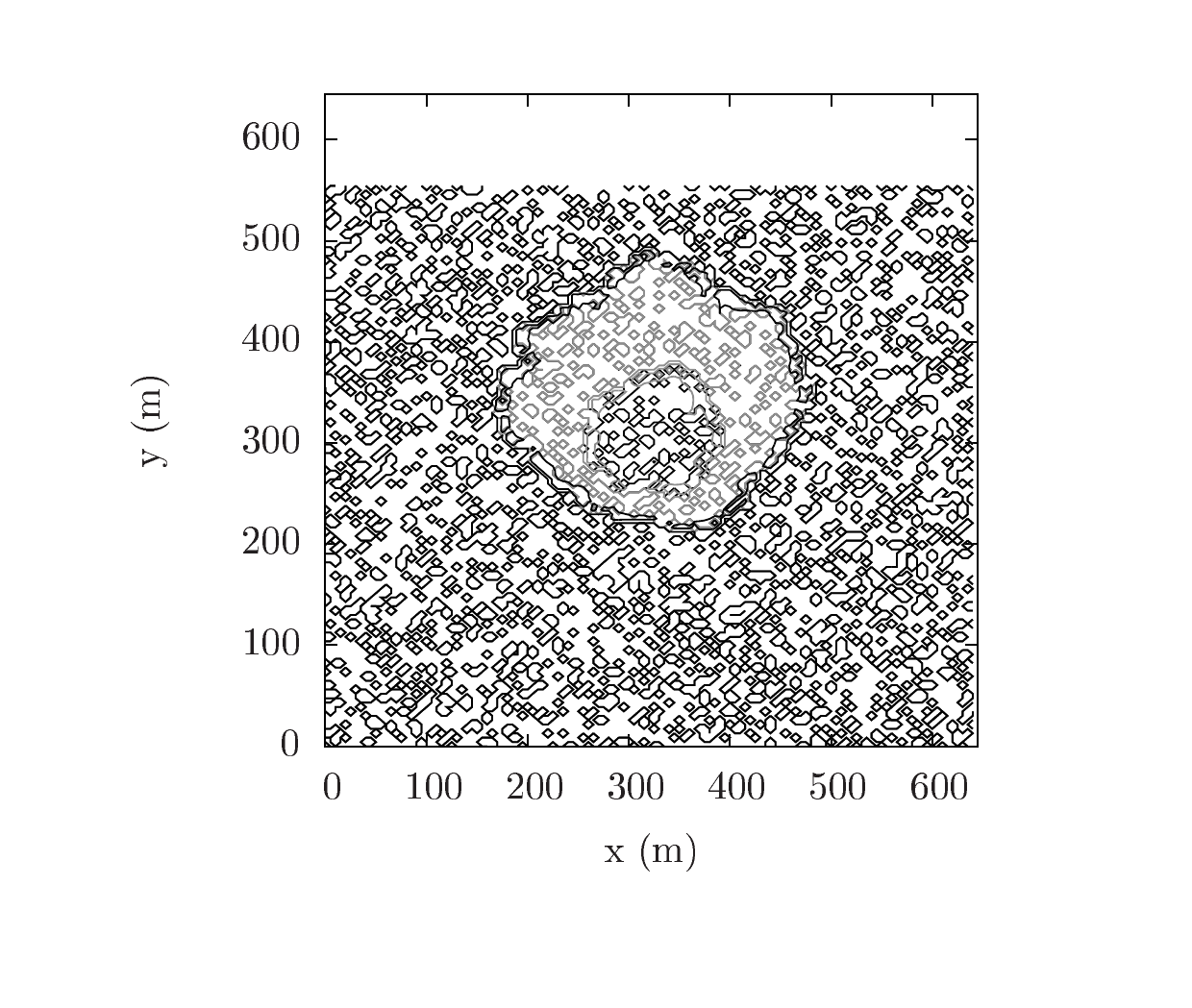}}\\
(b) \\
\end{tabular}
\caption{
(a)Evolution of the fire front ignited in the middle of the mesh, 
in a flat surface without slope and a dopping of $80\%$.
(b)In this case the surface has an inclination of $\pi/6\;[rad]$.
{\color{black} The alternated black and gray patterns represents the evolution of 
the system with a time difference of 10 minutes.}
\label{fig:estado}}
\end{center}
\end{figure}

{\color{black} We now} illustrate the spreading of fire through a bigger 
domain.
We use a square mesh of 129 sites per side, with a separation of 
{\color{black} 10} $[m]$ between each site.
In Fig. \ref{fig:estado} we show the time evolution of the fire 
front in a flat 
surface with no slope and other with an inclination of 
$\pi/6\;[rad]$, both with
an occupancy of the $80\%$.
The state of the system is obtained at different times, 
separated by intervals of approximately $5$ minutes.

The simulations {\color{black} produce} fire patterns accordingly to the expected 
behavior, 
climbing faster up hill and with just small effects of the 
underlying square mesh.

\subsection{Critical Behavior}

The last important point we will investigate, is the analysis of the
second order phase transition usually found in this kind of
systems. As before, the fire was simulated over a square lattice,
using a simple scenario of a flat surface with different inclinations,
ranging from $0$ to $\pi/3$, keeping the total distance between
neighboring sites fixed at $5\;[m]$ as before.  The runs starts with a
single site as a seed in the center of the mesh and are stopped when
the fire front reaches the borders of the domain or when the
propagation of the fire is stopped.  In this way the finite size
effects are irrelevant.  To keep the simulations as simple as
possible, we choose the neighborhood to be the eight closest sites to
each active site.  In this way the blocking effects are unimportant
and we can focus on the features of the spreading model (see
Eq. \ref{eq:dc-final}).

Although simple, the network spawned in this way brings some
theoretical complications, as we are no longer dealing with a simple
symmetric network, due to the weighting procedure explained in the
previous sections. For the low slope case, this will not be important,
but as the inclination grows this effect will be apparent and will not
allow us to use some of the usual constructs. In the following we will
try to keep the discussion in the more general way we can, avoiding
the use of quantities not always well defined.

To study criticality, 
{\color{black} we measure} the survival probability $P_{s}(t)$, 
the number of active sites $N(t)$
and the mean square radius, $R^{2}(t)$, of the infected zone 
{\color{black} starting from the ignition point}.
Those quantities are averaged over all surviving
runs. The literature shows that in this case it is reasonable to expect a behavior 
at the critical point as 
\cite{Dam03}:

\begin{equation}
P_{s}(t)\sim t^{-\delta},\; N(t)\sim t^{\theta},\;
R^{2}(t)\sim t^{2/z}
\end{equation}

The dynamical study of the critical behavior best suits this
particular configuration, since other quantities like correlation
length or cluster size loses their meaning in a non symmetric network,
as mentioned before.

\begin{figure}[!ht]
\begin{center}
\begin{tabular}[t]{c}
\resizebox{6.7cm}{!}{\includegraphics{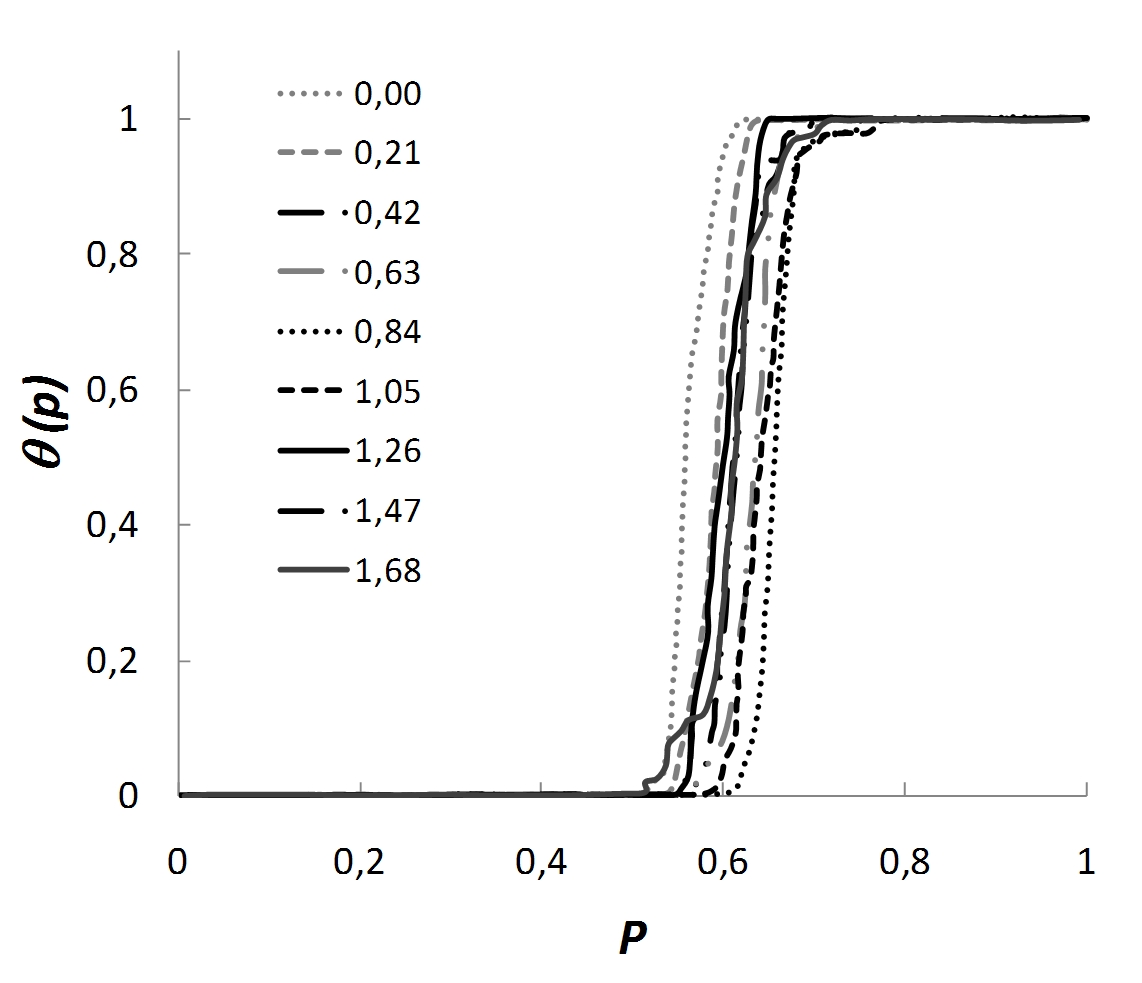}} 	\\
(a)  \\
\resizebox{6.7cm}{!}{\includegraphics{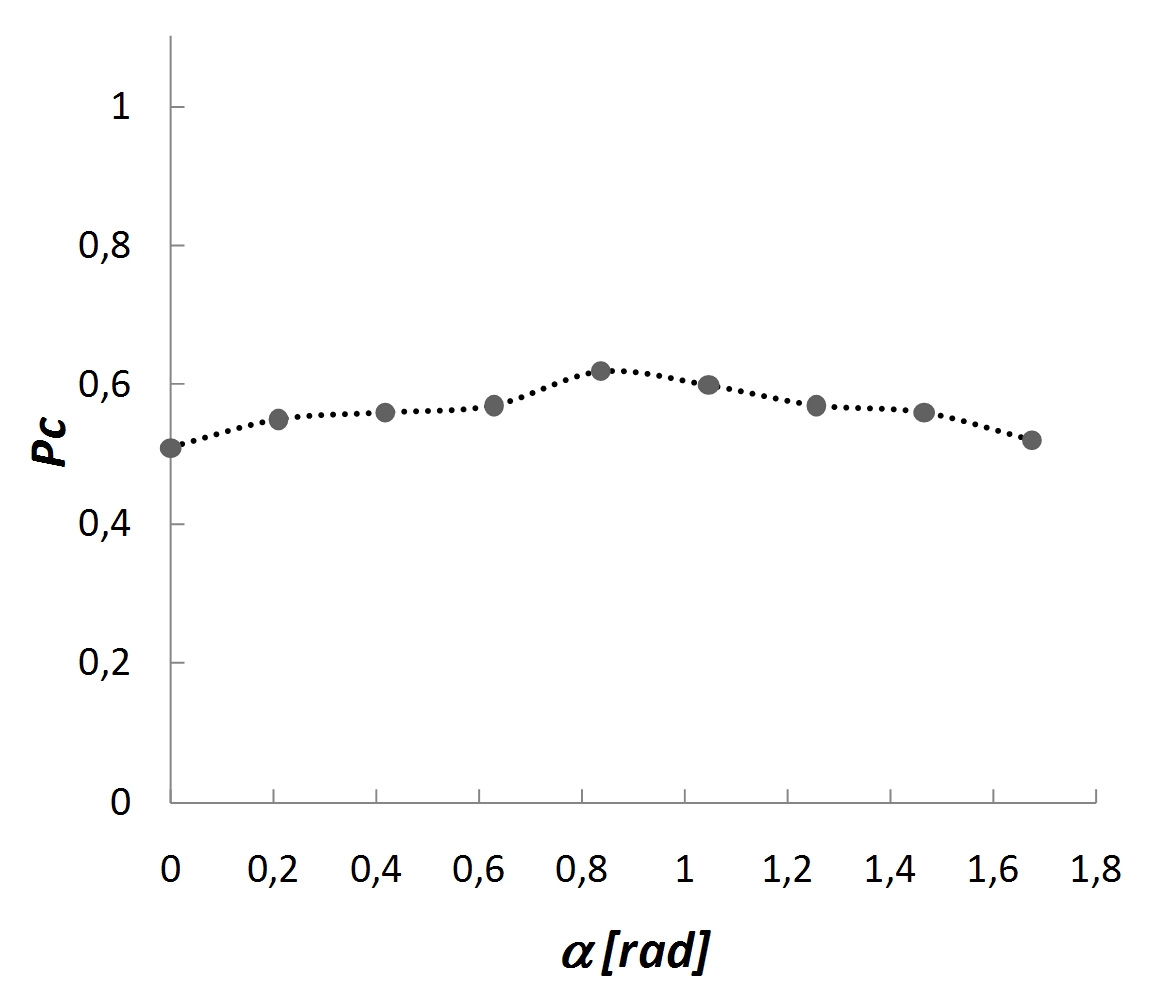}}\\
(b) \\
\end{tabular}
\caption{(a) The percolation density for different inclinations. 
(b) The percolation threshold as function of the slope.
\label{fig:theta}}
\end{center}
\end{figure}

In Fig. \ref{fig:theta}a the percolation probability $\theta(p)$, for
the different slopes is shown. Low inclinations in the burning surface
produces roughly constant percolation thresholds, reaching a peak
close to $0.8\;[rad]$ from where it decreases almost linearly for
higher slopes, as shown in Fig. \ref{fig:theta}b.

\begin{figure}[!ht]
\begin{center}
\begin{tabular}[t]{c}
\resizebox{6.7cm}{!}{\includegraphics{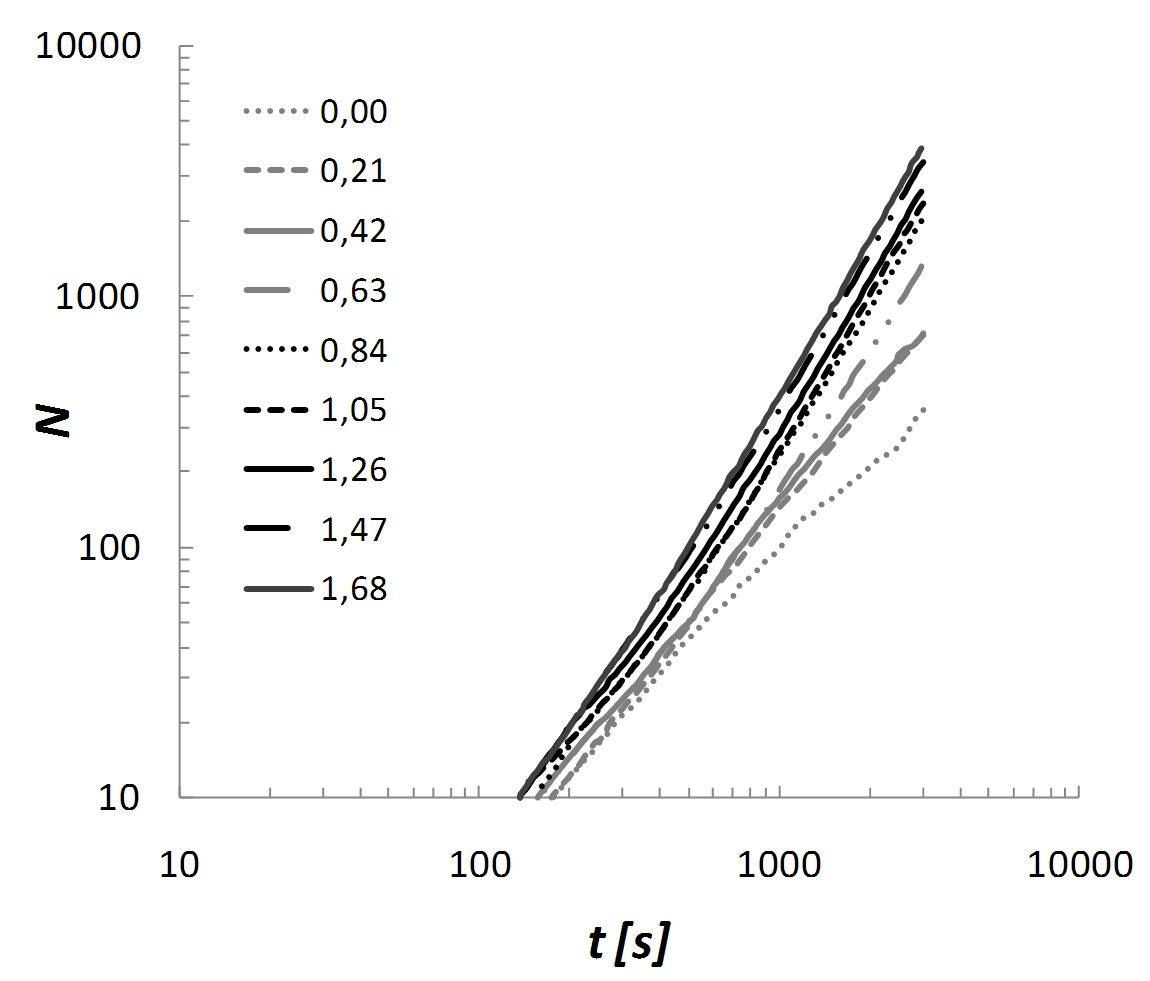}} 	\\
(a)  \\
\resizebox{6.7cm}{!}{\includegraphics{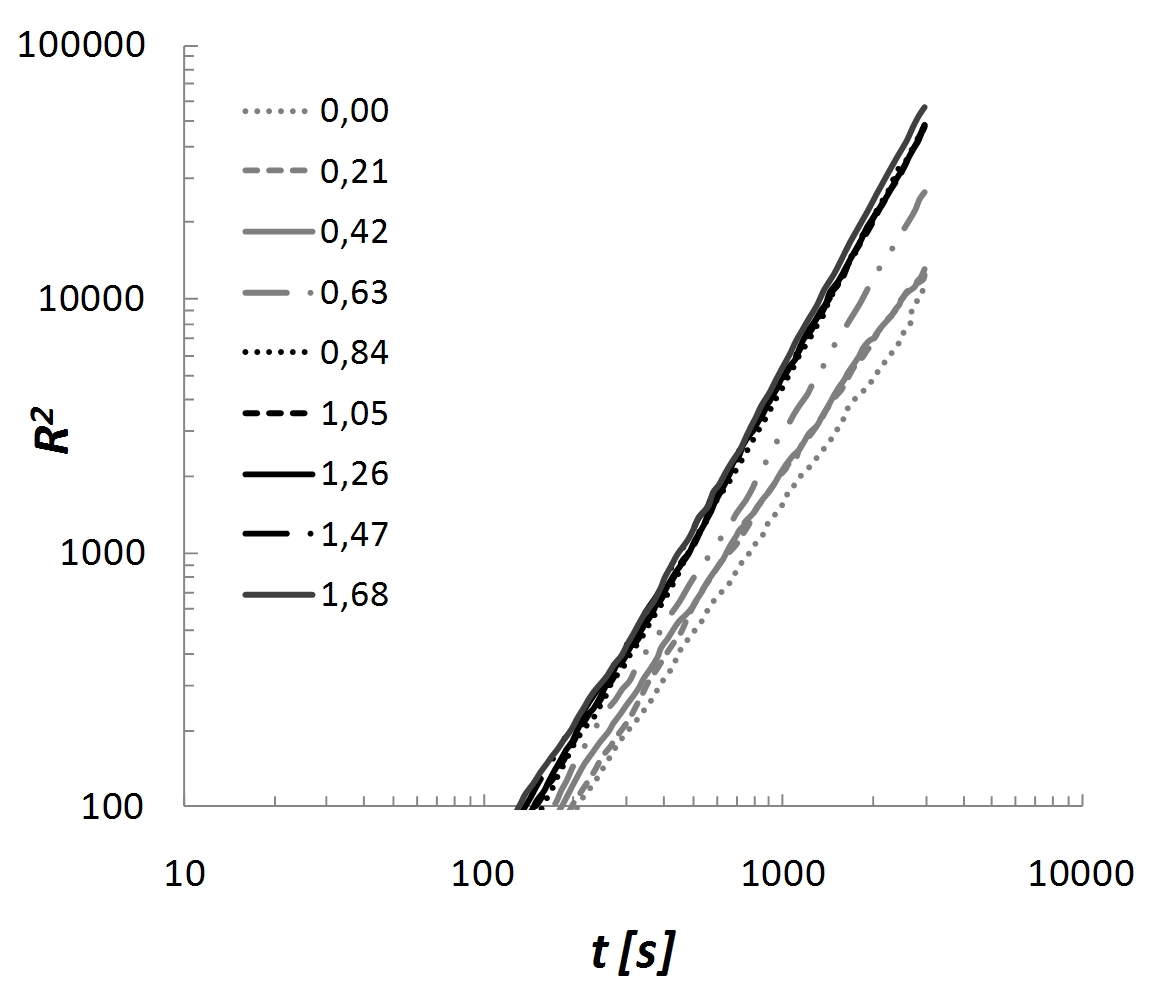}}\\
(b) \\
\end{tabular}
\caption{(a) The number of active sites $N(t)$. (b) The mean square radius R$^{2}$. Both results were obtained at criticality for different slopes $\alpha$. \label{fig:nr2}}
\end{center}
\end{figure}

At criticality, $N$ and $R^2$ are plotted in Fig. \ref{fig:nr2}a and
Fig. \ref{fig:nr2}b respectively. Both quantities {color{black} show}
a power law behavior at large times. The different inclinations,
depicted in different lines, produce different critical behaviors. In
both cases the critical exponent increases with the slope, thus
showing faster and more active fire spreads, as expected. For $N$ we
have that $\theta$ ranges from $1.173$ to approximately $1.996$,
whereas for $R^2$, ${2}/{z}$ has a minimum at about $1.705$ to a
maximum of $2.098$. Those ranges {\color{black} comprehend} the values
reported in \cite{Lin08} for the General Epidemic Process (GEP) in a
two dimensional lattice.

\begin{figure}[h!]
\begin{center}
\resizebox{20.0cm}{!}{\includegraphics*{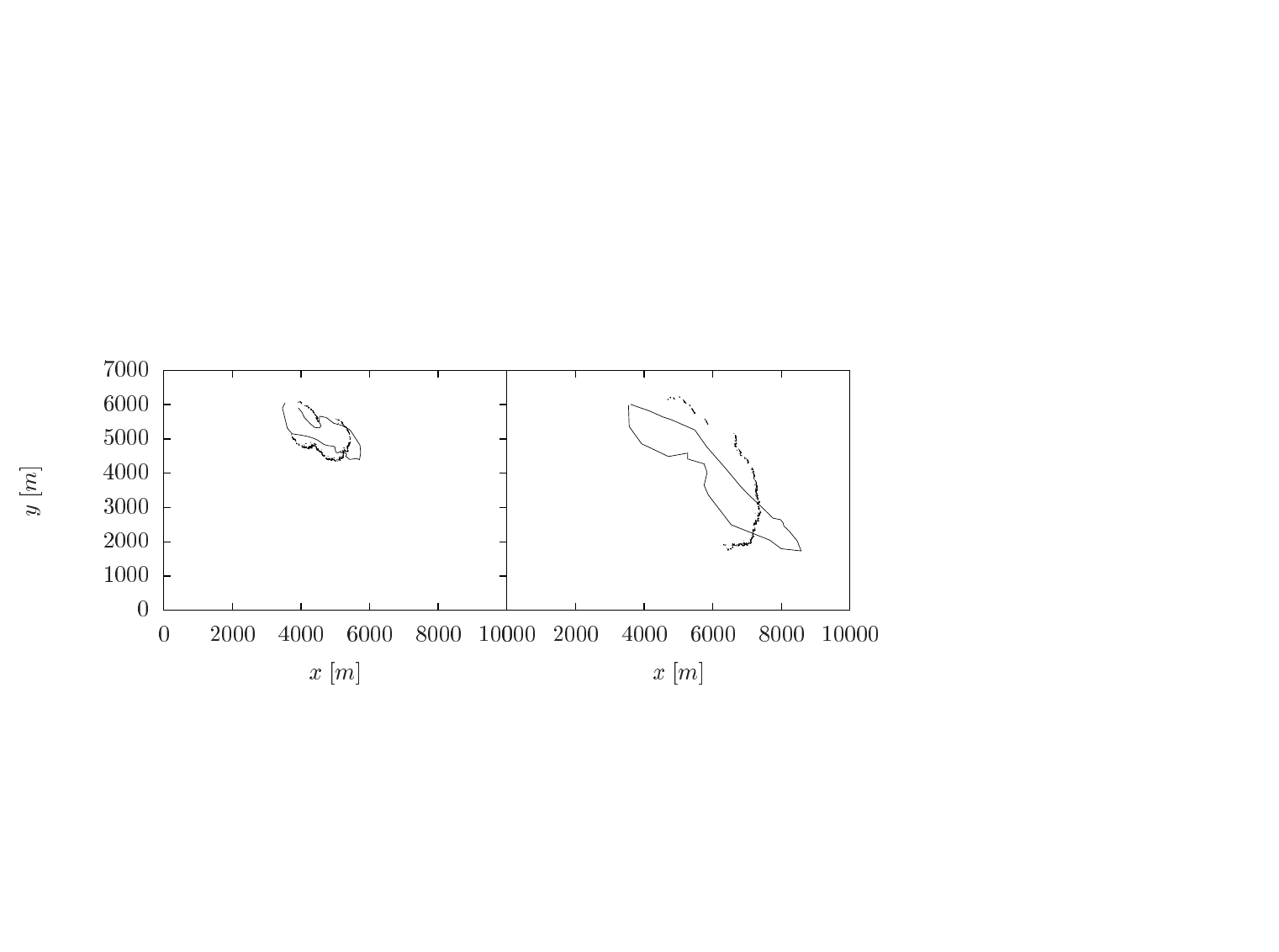}}
\caption{\label{fig:lancon} Fire front contours for the fire of Lan{\c c}on. 
The lines delimit the actual burned area, the points represent the simulated
fire front.}
\end{center}
\end{figure}

{\color{black}
\section{Experimental Validation}

In this last section we will compare our model to a forest fire
ignited near Lan{\c c}on in Provence, France, at 9:40 on July the
first of 2005.  There were reported two ignition points and the mean
wind speed reached $46\;[km/h]$ at $10\;[m]$ avobe the ground, with an
average wind direction of about $330 \deg$ (NNW).

The main vegetation type present in the fire is the kermes oak shrubs
(Quercus coccifera), for which we use a mean life of $17\;[s]$ and a
radiation constant $\alpha=4.2\;10^{-6}\;[J^{-1}]$. The vegetation and
terrain data is available at a resolution of $50\;[m]$, so for the
terrain we use a bilinear interpolation and for the vegetation, the
$50\;[m]$ sided cell is filled with kermes oaks at a distance of
$3.7\;[m]$ with an ocupancy probability of $0.64$.  For more details
of this scenario see {\it e.g.} Ref.\cite{Ado10}.

Data of the fire front advance is available at 12:00 and 14:34 hrs.
In table \ref{tab:lancon} we present the burned area and the rate of
spread (ROS) of the fire and the simulation results, and in
Figure \ref{fig:lancon} an aerial comparisson of the numerical results
versus the data.

\begin{table}
\begin{tabular}{llcc}
                   &Time &Fire &Simulation\\
\hline \\
ros  (m/h)         & 12:00& 714 & 1\\
                   & 14:34& 1523& 1\\
Burned Surface (ha)& 12:00& 148 & 1\\ 
                   & 14:34& 485 & 1\\ 
\end{tabular}
\caption{Rate of spread and burned surface for the fire of Lan{\c c}on and simulation. }
\label{tab:lancon}
\end{table}

The burned area is slightly overestimated. Part of this discrepancy
can be explained by the action of firefighters starting around 13:00
hrs. The model gives a reasonable description of the fire pattern,
inspite of the roughness of the radiative model and the physical
description of the vegetation content. Work in this respect is being
carried out \cite{Bil10} and we envisage a future improvement on those
areas. None the less, the propagation model \ref{eq:dc-final} shows
the relevant features one expects and seems to be adecuate for
real-life scenarios.

}

\section{Conclusions} 

We have developed and tested a new continuous model for the fire
spread under the $lswn$ approach. The model is controlled by a
phenomenological dynamics and is able to reproduce realistic spreading
scenarios. It allows including heterogeneous vegetation (type,
geometry, mass, etc.), as well as changes in the topography, this
latter characteristic being one of interest in the present work. The
heterogeneity of the landscape is captured via geometric factors
included locally in each site as parameters of the dynamic
equations. This gives a simple starting point to future studies where
more effects will be included. For example, the wind effects can be
introduced in the description modifying only the geometric factors
between each site.

In the present study, the model was used to simulate the fire
spreading process in different landscapes for a defined kind of
vegetation. First, it was shown that the fire spread exhibits--as the
classical percolation picture--a continuous phase transition between a
self-extinguishing state and an infinite spreading one, depending on
the occupancy probability $p$. Secondly, it was observed that for
surfaces with constant slope, the percolation threshold shows an
increasing behavior with the inclination, which produces greater
critical exponents for inclined configurations.

We plan the introduction in this model of a more complete description
of the sites network, including for example an amorphous mesh over a
fractal landscape and where the blocking of trees are fully taken into
account.

\section*{Acknowledgments} 
This work was funded by SCAT-ALFA project and supported by the CNRS (ANR PIF/NT05-2 44411 and GDR 2864).
\newpage

\newpage
\listoffigures


\newpage

\small
\baselineskip 10pt

\newpage
\newpage
\newpage
\newpage
\newpage
\newpage

\end{document}